\documentclass[10pt, conference, compsocconf]{IEEEtran}

\usepackage[noadjust,nospace]{cite}
\usepackage{booktabs}
\usepackage{amsfonts}
\usepackage{amsmath}
\usepackage{xspace}
\usepackage{subcaption}
\usepackage{tikz}
\usetikzlibrary{automata, positioning}
\usepackage{multirow}
\usepackage{amssymb}

\usepackage{comment}

\definecolor{DMlightblue}{RGB}{166,206,227}
\definecolor{DMdarkblue}{RGB}{31,120,180}
\definecolor{DMlightgreen}{RGB}{178,223,138}
\definecolor{DMdarkgreen}{RGB}{51,160,44}
\definecolor{DMlightred}{RGB}{251,154,153}
\definecolor{DMdarkred}{RGB}{227,26,28}
\definecolor{DMlightorange}{RGB}{253,191,111}
\definecolor{DMdarkorange}{RGB}{255,127,0}
\definecolor{DMpastel}{RGB}{202,178,214}

\begin{document}

\author{\IEEEauthorblockN{Valentin Poirot}
\IEEEauthorblockA{\textit{Kiel University}, Germany \&\\
\textit{Chalmers University of Technology}, Sweden \\
vpo@informatik.uni-kiel.de}
\and
\IEEEauthorblockN{Olaf Landsiedel}
\IEEEauthorblockA{\textit{Kiel University}, Germany \&\\
\textit{Chalmers University of Technology}, Sweden \\
ol@informatik.uni-kiel.de}
}

\title{Dimmer: Self-Adaptive Network-Wide Flooding with Reinforcement Learning}

\maketitle
\begin{abstract}
The last decade saw an emergence of Synchronous Transmissions (ST) as an effective communication paradigm in low-power wireless networks.
Numerous ST protocols provide high reliability and energy efficiency in normal wireless conditions, for a large variety of traffic requirements.
Recently, with the EWSN dependability competitions, the community pushed ST to harsher and highly-interfered environments, improving upon classical ST protocols through the use of custom rules, hand-tailored parameters, and additional retransmissions.
The results are sophisticated protocols, that require prior expert knowledge and extensive testing, often tuned for a specific deployment and envisioned scenario.
In this paper, we explore how ST protocols can benefit from \textit{self-adaptivity}; a self-adaptive ST protocol selects itself its best parameters to (1) tackle external environment dynamics and (2) adapt to its topology over time.

We introduce \textit{Dimmer} as a self-adaptive ST protocol.
Dimmer builds on LWB and uses Reinforcement Learning to tune its parameters and match the current properties of the wireless medium.
By learning how to behave from an unlabeled dataset, Dimmer adapts to different interference types and patterns, and is able to tackle previously unseen interference.
With Dimmer, we explore how to efficiently design AI-based systems for constrained devices, and outline the benefits and downfalls of AI-based low-power networking.
We evaluate our protocol on two deployments of
resource-constrained nodes
achieving 95.8\% reliability against strong, unknown WiFi interference.
Our results outperform baselines such as non-adaptive ST protocols ($\sim$27\%) and PID controllers, and show a performance close to hand-crafted and more sophisticated solutions, such as Crystal ($\sim$99\%). 
\end{abstract}

\begin{IEEEkeywords}
low-power wireless networks, synchronous transmissions, reinforcement learning, deep Q-network, WSN, IoT
\end{IEEEkeywords}
\section{Introduction}
\label{sec:introduction}

Energy-efficiency and high-reliability are two vital aspects of low-power wireless communication.
Synchronous Transmissions~(ST), with Glossy as their flagship~\cite{Ferrari2011-Glossy}, provide high performance under normal conditions, i.e., under no or minimal interference~\cite{Ferrari2011-Glossy,Ferrari2012-LWB,AlNahas2017-A2}; yet, the wireless medium is prone to large dynamics, e.g., due to fading and interference.
With the EWSN competitions, the community has designed dependable ST protocols, able to maintain communication under strong interference~\cite{Lim2017-RobustGlossy, Istomin2018-Crystal2, Ma2019-DeCOT}.
Through cleverly-crafted rules and heavy testing, dependable solutions provide high-reliability during highly-interfered episodes, often at the cost of lower energy-efficiency.
Beyond dependability, we argue that the next step towards generalized ST protocols is to provide \textit{adaptivity};
an adaptive protocol detects changes to its environment, e.g., interference, and reacts to counteract its effects, e.g., by updating its transmission strategy or excluding jammed frequencies.
In this paper, we explore \textit{self-adaptivity}; rather than relying on predefined and hand-crafted decision-rules, a self-adaptive ST protocol learns by itself how to detect interference and react to it.

Adaptivity plays an important role in communication engineering, and has been shown to improve performance, e.g., in TCP congestion control~\cite{RFC5348}, or WiFi rate control~\cite{WiFi}.
A self-adaptive wireless stack faces many challenges, as interference comes in many patterns (e.g., bursts or slow channel-fading)~\cite{Srinivasan2008-burstiness}, and is often unique to each deployment.
Further, device positions, density, and mobility, have a drastic impact on performance.
In ST protocols, increasing the number of retransmissions is shown to increase reliability under interference, along with using channel-hopping~\cite{Lim2017-RobustGlossy, Istomin2018-Crystal2}.
Yet, in dense ST deployments and in the absence of interference, not all concurrent transmissions are required to ensure correct reception~\cite{Carlson2013-XFCS,Sarkar2016-Sleeping-Beauty}.
Likewise, leaf nodes at the edge of a non-mobile network are not helpful to the dissemination~\cite{Zhang2017-LessIsMore}.
In such cases, devices do not need to participate in the flood propagation and can be deactivated earlier to save energy.
Thus, we define an ST protocol as adaptive if (1) it reacts to external environment dynamics and (2) adapts to its topology over time, and define it as self-adaptive if it is able to learn by itself how to be adaptive.

\textbf{Challenges.}
Designing a self-adaptive ST protocol brings the following challenges: available information is limited, as Glossy does not provide feedback, or feedback is delayed in shared-bus abstractions~\cite{Ferrari2012-LWB}.
Further, ST abstract away the concept of neighbors:
one-hop and multi-hop neighbors are indistinguishable.
Thus, interference must be dealt with globally, as distributed approaches might cause instability~\cite{Zhang2017-LessIsMore}.
Rule-based systems used in adaptive rate control~\cite{RFC5348} react to disturbances, but require complex rules for efficient or near-optimal adaptation.
PID controllers, the go-to approach in closed-loop control, must be tuned, either through experimentation or complex numerical methods, prior to their deployments~\cite{Ang2005-PID-tuning}; yet it is not guaranteed that different interference-patterns or deployments require similar decision-rules, thresholds, or parameters.
Instead of static-rules or PID controllers, we decide to employ deep Reinforcement Learning (RL) in a bid to learn how to optimally tune our communication parameters and react to deployment-specific interference, in the absence of both human supervision and expert knowledge.
Using deep RL, we achieve self-adaptivity: our self-adaptive protocol learns by itself how to detect and react to its environment dynamics.

\textbf{System Challenges.}
Building on neural networks and deep RL brings its own, specific set of challenges:
(1) We have to capture the dynamics of the wireless medium in a neural network so that it can learn how to efficiently adapt to the network dynamics.
We use the packet reception rate and the radio duty-cycle as measures of interference.
(2) It is practically impossible to obtain a dataset labeled with optimal parameters, as used with supervised methods.
We build an unlabelled simulation environment in which a reinforcement-learning agent is trained.
(3) Low-power wireless systems are resource-constrained, in the order of several MHz and tens of kB of RAM,  and demand for space-efficient neural networks so that we can deploy them on the hardware.
We employ quantization and a small architecture to limit space.
(4) Distributed RL approaches are prone to instability, while building a fully central RL solution is infeasible due to the curse of dimensionality.
We employ a central deep network to globally adapt to interference, and distributed multi-armed bandits to locally deactivate devices.

\textbf{Approach.}
We introduce \textit{Dimmer}, a self-adaptive ST protocol for all-to-all communication.
Dimmer is part of the Low-power Wireless Bus (LWB) class of protocols~\cite{Ferrari2012-LWB}: applications see the medium as a shared bus, and a coordinator centrally schedules communication into periodic rounds.
Dimmer introduces two novel elements: (1) a centrally-executed Deep Q-Network (DQN) that globally adapts the retransmission parameter to tackle interference, and
(2) a distributed forwarder selection scheme using multi-armed bandits at runtime, to deactivate superfluous devices and save energy in the interference-free case.
Unlike other forwarder-selection approaches~\cite{Sarkar2016-Sleeping-Beauty}, Dimmer does not require extra transmissions:
application packets are enhanced with local performance measurements and shared with the network; and communication schedules are used to globally adapt devices to the current measured interference.
At the end of a round, the Dimmer coordinator aggregates the feedback received from all nodes and executes its DQN, thus establishing a new transmission strategy for the entire network.
In the interference-free case, the coordinator allows devices to sequentially learn at runtime if they are essential to the information dissemination or can be deactivated to save energy.

\textbf{Contributions.}
This paper contributes the following:
\begin{itemize}
    \setlength\itemsep{0em}
    \item We present Dimmer, an RL-enabled self-adaptive communication primitive featuring a deep Q-network and adversarial multi-armed bandits;
    \item We highlight how we represent Dimmer as a solvable RL problem featuring centralized control and distributed decision-making, and how we design an embedded deep Q-network fitting to low-power platforms;
    \item We provide an open-source\footnote{https://github.com/ds-kiel/dimmer} implementation and evaluate our solution on two testbeds comprising 18  and 48 nodes, showing it is able to operate on new topologies and adapt against unseen interference without retraining its DQN.
\end{itemize}

We give background on RL and LWB in~\S\ref{sec:background}, provide an overview of Dimmer in ~\S\ref{sec:overview}, dive deep into our problem formulation and system design in \S\ref{sec:design}, evaluate Dimmer on testbed deployments in~\S\ref{sec:evaluation}, discuss related work in~\S\ref{sec:related_work} and conclude our work in~\S\ref{sec:conclusion}.

\section{Background}
\label{sec:background}


\subsection{Reinforcement Learning}
\label{sec:background:RL}

Reinforcement Learning (RL) targets \textit{sequential decision making}~\cite{Sutton1998};
in an RL problem, an agent learns how to achieve a complex task by taking consecutive actions in an environment with unknown rules and dynamics.
%
Learning happens through active interaction:
the agent measures its environment, then executes actions affecting it back;
a reward signal indicates if the desired goal is reached.
By exploring the environment and trying random actions, the agent accumulates experiences and builds an internal model.
By exploiting its reward and with enough knowledge, the agent constructs a sequence of actions that solves the complex task at hand.

\textbf{Markov Decision Processes.}
An RL problem is said to be solvable if the environment can be represented as a Markov Decision Process (MDP).
MDPs extend Markov chains: 
the transition to a new state is modeled as a transition probability-distribution, from a state-decision pair.
Formally, an MDP is represented by the tuple $(\mathcal{S},\mathcal{A},\mathcal{P},\mathcal{R})$, where $\mathcal{S}$ is the set of states, $\mathcal{A}$ the set of actions, $\mathcal{P}$ the transition probability-function and $\mathcal{R}$ a reward function.

\textbf{Q-Learning.}
An established way to solve an RL problem is to maximize the cumulative reward $R_t \triangleq \sum_{\tau=t}^{\infty}{\gamma^{\tau-t}r_{\tau}}$, where $r_{\tau}$ is the reward obtained when transitioning at time $\tau$, and $\gamma \in [0,1)$ a constant called the discount factor.
Small discount factors force the agent to maximize short-term, immediate rewards, while high discount factors allow the agent to maximize long-term expected rewards.
Q-learning is one of the most popular ways to solve RL problems \cite{Sutton1998}.
In Q-learning, an agent learns an action-value function $Q(s,a)$.
This function represents the expected cumulative reward the agent expects when starting in state $s$, using action $a$ such as $Q(s,a) \triangleq\nolinebreak \mathbb{E}\big[ R_t \mid s_t =\nolinebreak s, a_t =\nolinebreak a\big]$.
In simple terms, the Q-function evaluates how valuable it is to choose action $a$ in state $s$, in terms of expected reward.

If the environment can be modeled as an MDP, then we can find an optimal function $Q^{*}(s,a)$ that follows Bellman's principle of optimality:
\begin{equation}
   \label{eq:qfunction_bellman}
    Q^{*}(s,a) = \mathbb{E}\big[ r_{t} + \gamma{}  \textrm{max}_{a'}Q^{*}{(s_{t+1},a')}\mid s_t = s, a_t = a\big]
\end{equation}
where $r_{t}$ is the immediate reward received, $\gamma$ the discount factor, and $s_{t+1}$ the state achieved after the state $s$.
By iteratively trying actions and receiving rewards, we can update a Q-function that ultimately converges to the optimal $Q^{*}(s,a)$.

\textbf{Deep Q-learning.}
While Q-learning algorithms have historically used a tabular approach to represent the Q-function \cite{Sutton1998}, deep neural networks have been recently established as Q-function approximators \cite{Mnih2015-DQN}.
Deep Q-networks (DQN) have been successfully used to solve problems outside of their original application: datacenter cooling \cite{Lazic2018-datacenter}, wireless modulation \cite{Vrieze2018-RLModul}, CSMA/CA optimization \cite{Mastronarde2016-RL-CSMA}, etc.
One advantage of DQN over tabular approaches is the ability to solve problems with continuous states and the generalization property of neural networks.

\textbf{Multi-armed bandits.}
In the Multi-Armed Bandits (MAB) problem, a gambler faces $K$ machine slots in a casino, each machine giving an a priori unknown, stochastic reward upon pulling its arm. 
The goal is to maximize the cumulative returns in a minimal number of steps, by carefully controlling the exploration-exploitation trade-off.
In the extended adversarial MAB setting, an adversary is able to impact the reward system associated with each arm~\cite{Auer1995-adversarial-mab}.
In wireless systems, changing conditions of the medium can be represented as an adversarial setting.
The gambler thus cannot rely on past experiences only, and must continuously explore.

\textbf{Exp3.}
Exp3 is an online learning approach for adversarial MAB~\cite{Auer2002-exp3}.
Exp3 associates an exponential weight with each arm, thus leading to quick adaptation to adversarial changes in the environment.
At each timestep, an action is selected based on its probability, such as:
\begin{equation}
   \label{eq:exp3}
    p_i(t) = (1-\gamma)\times\frac{w_i(t)}{\sum_{j=1}^K{w_j(t)}}+\frac{\gamma}{K}
\end{equation}
Where $w_i(t)$ is a weight updated after each trial such as $w_i(t+1) =\nolinebreak w_i(t) \times exp(\frac{\gamma*r(t)}{K*p_i(t)})$, with $r(t)$ the reward obtained at time $t$.
$K$ stands for the number of arms, and $\gamma$ is the exploration factor.

\subsection{Synchronous Transmissions}
\label{sec:background:LWB}

In low-power wireless networks, quickly flooding a message to the entire network has established itself as a simple and efficient method to provide communication.

\textbf{Glossy.}
Glossy is among the pioneer works in synchronous transmissions \cite{Ferrari2011-Glossy}.
Through tight synchronization ($<$~0.5~$\mu$s) and by sending identical data, Glossy provides network-wide broadcast with high reliability ($>$~99.9\%) and low power consumption.
Within a Glossy flood, a packet is retransmitted multiple times, typically 3, and nodes alternate between transmission and reception to keep synchronization and reduce the number of concurrent transmissions.
Variants extend Glossy with flexible transmission schedules and frequency hopping~\cite{Lim2017-RobustGlossy}.

\textbf{LWB.}
Low-power Wireless Bus (LWB) is a flexible communication protocol, supporting many traffic patterns and specially tailored to wireless sensor networks \cite{Ferrari2012-LWB}.
LWB uses Glossy floods as communication primitive, effectively turning a multi-hop network with mobile nodes into a logical bus, in which any node can potentially receive any packet, without the need for expensive routing.
LWB is a centralized solution:
a host node computes a schedule that satisfies flows requested by (message)-source nodes and controls the periodicity of communication to save energy.
As such, LWB is a versatile solution for low-power communication, and has inspired higher-level abstractions in low-power wireless systems~\cite{Ferrari2013-VIRTUS, Jacob2019-Baloo}.
\section{An Overview of Dimmer}
\label{sec:overview}

\begin{figure*}[tb]
    \centering
    \includegraphics[width=0.8\textwidth]{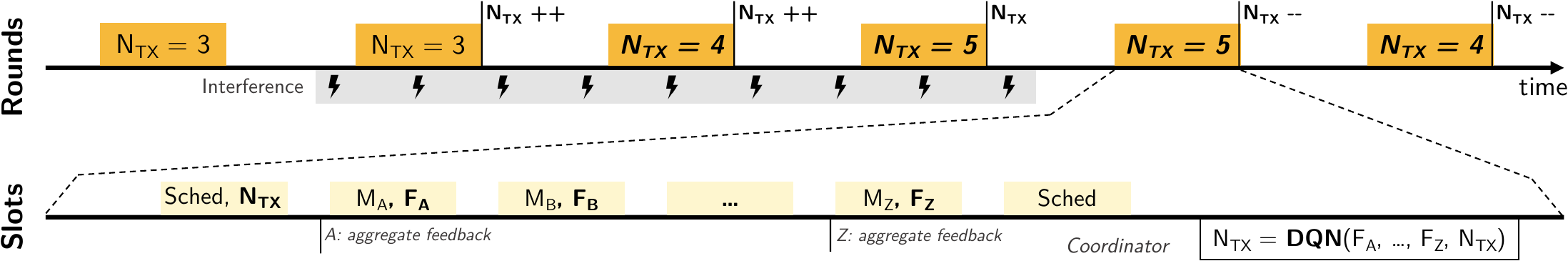}
    \caption{Adaptivity in Dimmer. A round starts with a Glossy flood from the coordinator containing the round schedule and the retransmission parameter $N_{TX}$. Each device sends its message $M_x$ and its performance feedback $F_x$. At the end of the round, the coordinator aggregates the feedback and computes the new $N_{TX}$ with its Deep Q-Network. The retransmission parameter $N_{TX}$ increases to counteract interference, and converges back to its optimal value once interference has passed.}
    \label{fig:central_adaptivity}
\end{figure*}
\begin{figure}[tb]
    \centering
    \includegraphics[width=0.75\columnwidth]{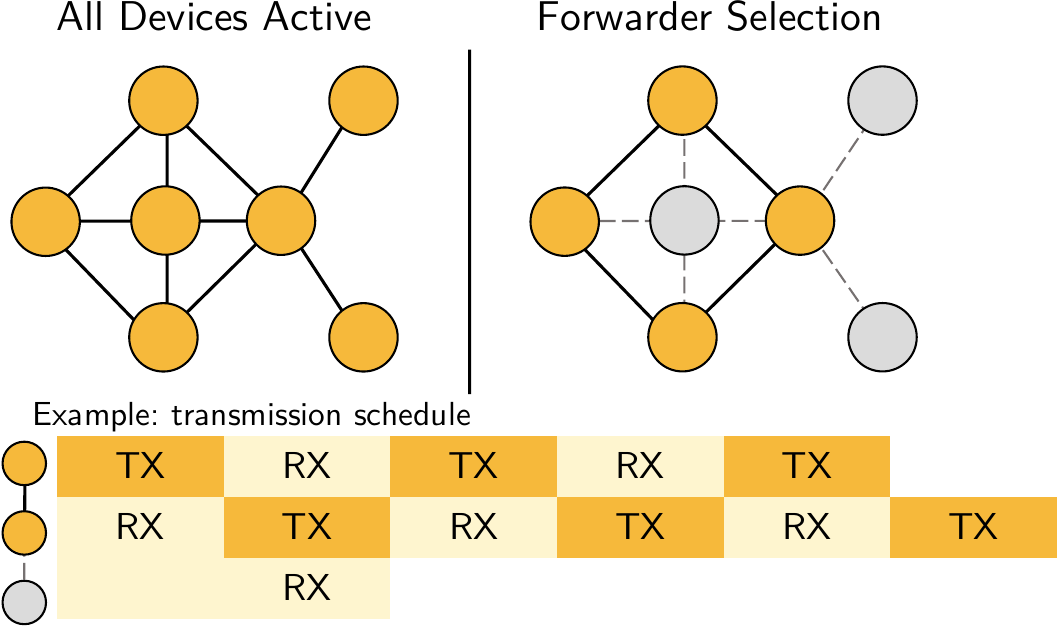}
    \caption{Forwarder Selection. In the interference-free case, nodes take turn learning whether they should act as active forwarder to help the dissemination, or passive receiver to save energy. Under interference, all devices are active.}
    \label{fig:forwarder_selection}
\end{figure}

\subsection{Dimmer}

We introduce Dimmer, a self-adaptive synchronous-transmissions (ST) protocol.
Dimmer provides self-adaptivity: through deep reinforcement learning and without expert supervision, Dimmer learns by itself how to detect changes to the wireless medium, and reacts by updating its retransmission strategy to counteract losses or save energy.

\textbf{Dimmer in a nutshell.}
Dimmer uses LWB's communication structure: a central coordinator schedules periodic communication rounds (see Fig.~\ref{fig:central_adaptivity}).
A round starts with a control slot, used to transmit the schedule.
Then, data slots are attributed to each device to allow message dissemination, where each slot is executed as a Glossy flood.
In Dimmer, each device continuously monitors its performance, i.e., its local packet reception rate and average radio-on time.
During its data slot, a source includes its performance in a two-byte packet header, and shares it with the network.
Devices collect such feedback from all participating nodes.

At the end of the round, the central coordinator executes its embedded, deep Q-network over the collected feedback, and decides a new retransmission strategy.
This updated strategy is shared along with the schedule, at the beginning of the next round, and is applied by the entire network.
If no interference is detected, the central coordinator instead allows devices to apply their distributed forwarder selection, where one node learns whether its participation is required for successful message propagation, see Fig.~\ref{fig:forwarder_selection} and \S\ref{sec:design:MAB}.

\subsection{AI versus traditional methods}

\textbf{Limitations of traditional methods.}
Building an adaptive ST protocol using rate-control rules or PID controllers is possible but suffers drawbacks.
Rule-based rate controllers often provide adaptivity by overshooting the optimal value: in TCP, the data rate is greatly decreased whenever it suffers losses.
Similarly, PID controllers must be tuned, either through numerical analysis or experimentation.
Thus, efficient, traditional adaptivity requires sufficient expert knowledge and extensive testing; yet it is not guaranteed whether the tuned parameters, complex rule-sets and chosen thresholds are optimal in unknown deployments.

\textbf{AI-enabled wireless.}
With machine learning, deploying to a different topology or changing the hardware simply equates to collecting new traces and retraining the neural network, which can be done as an automatic step during deployment.
Moreover, RL does not require a dataset labeled with the optimal retransmission parameters, as the RL agent uses trial-and-error to learn how to act optimally.
Thus, using Dimmer does not require any prior expert knowledge.
In addition, we show in \S\ref{sec:evaluation} that deploying Dimmer in a new environment does not necessarily require retraining the DQN, as we demonstrate by operating Dimmer on a larger deployment against WiFi, while the DQN was trained on traces collected on an 18-node testbed predominantly  featuring IEEE~802.15.4 jamming.

\textbf{Deep RL.}
Traditional, tabular Q-learning provides learning with low-complexity costs, yet only supports problems with low-dimensional states.
In Dimmer, our input space is a combination of measures from many nodes, particularly energy, that is continuous in nature.
This high-dimensionality makes tabular Q-learning unfit.
Rather, we rely on neural networks and deep RL.

\section{Problem Formulation and Design}
\label{sec:design}

We formulate the problem of adaptivity in \S\ref{sec:design:adaptivity}, and frame it as two independent RL problems in \S\ref{sec:design:DQN} and \S\ref{sec:design:MAB}.
We summarize our architecture in \S\ref{sec:design:architecture} and discuss selected challenges in \S\ref{sec:design:discussion}.

\subsection{Adaptivity: Two Sub-problems}
\label{sec:design:adaptivity}

To be adaptive, Dimmer needs to (1) counteract external environment dynamics and (2) deactivate superfluous concurrent transmitters and leaf devices in the interference-free case.
We argue that the number of retransmission within a Glossy flood, $N_{TX}$, is a parameter allowing to improve resiliency to external disturbances.
Further, setting $N_{TX}=0$ allows a participant to turn off its radio as soon as a packet is received once, thus saving energy.
We conclude that Dimmer is adaptive if it assigns (possibly unique) $N_{TX}$ values to each device in the network.

\textbf{Curse of dimensionality.}
We argue that it is practically hard to design a learning solution to adaptivity that is either fully centralized, or fully distributed.
Assuming a device chooses from $N_{TX}=0$ to $N_{max}$, and a network composed of $D$ devices, there are $(N_{max}+1)^D$ possible configurations, where many configurations actually break communication.
To find the optimal configuration, a learning system would require an enormous amount of data; this is not practical.

Let's instead assume a fully distributed adaptive system.
Each device selects its own $N_{TX}$.
If all participants try out strategies concurrently, it is practically impossible to discern whether a local strategy has been beneficial or if a concurrent strategy has been harmful.
If devices learn in a sequential manner, the time required increases linearly with the number of devices, and collecting traces to speed up the process is not practical, for the same reason as outlined for the centralized approach.
Further, distributed learning approaches are not guaranteed to converge to the optimal configuration~\cite{Zhang2017-LessIsMore}.

\textbf{Two sub-problems.}
To solve the challenge discussed above, we divide Dimmer into two distinct sub-problems:
(a) a centralized adaptivity control, that leads a global update of $N_{TX}$ under interference (Fig.~\ref{fig:central_adaptivity}), and (b) a distributed forwarder selection scheme, that allows devices to learn if their participation in the forwarding process is required in non-interfered episodes (Fig.~\ref{fig:forwarder_selection}).
Our centralized adaptivity leads to $N_{max}+1$ global configurations, assuming $N_{TX}=0$ to $N_{max}$, as all devices now share a common value.
Therefore, it is feasible to collect traces for such a problem, under varied disturbances, and learn an optimal solution (see \S\ref{sec:design:DQN}).

By making the distributed forwarder selection problem a binary choice (active forwarder, passive receiver), we have $2^D$ possible configurations.
It is not possible to further simplify the problem: $D$ is the number of devices in the entire network, and we cannot cluster devices into smaller, independent groups.
Indeed, different network subsets can depend on the decision of a distant, bottleneck device, thus finding independent clusters is practically hard.
Instead, we decide to tackle forwarder selection in a distributed manner.
Since collecting traces for such a number of configurations is practically unfeasible, Dimmer learns its forwarder selection configuration at runtime.
We allow devices to sequentially learn their role, one at a time, and measure their impact on the overall performance.
By learning sequentially, we ensure that, from a node perspective, the environment is stable.
We describe the techniques we employ to avoid network-breaking configurations and divergent states in \S\ref{sec:design:MAB}.

\subsection{Central Adaptivity}
\label{sec:design:DQN}

\textbf{Problem Formulation.}
We formulate the task of centralized adaptivity control in Dimmer as an RL problem.
\\
\textbf{Objective:} Find the global, optimal flood retransmission parameter $N_{TX}$ that maximizes reliability and minimizes energy consumption at a given time.
\\
\textbf{State space: } Dimmer aggregates the (a) \textit{reliability} (packet reception rate, ratio of packets received / expected) and (b) \textit{radio-on time} of a subset of the $K$ devices with lowest reliability.
In addition, the (c) \textit{current $N_{TX}$} is used, and represented as one-hot encoding.
(d) \textit{$M$ historical reliability datapoints} are used to encode past events, see Table~\ref{tab:features}.
\\
\textbf{Action space:} Dimmer applies to all nodes in the network the same action: (a) Decrease $N_{TX}$, (b) Maintain $N_{TX}$, or (c) Increase $N_{TX}$.
\\
\textbf{Neural architecture:} Dimmer uses one fully-connected hidden layer of 30 neurons with rectified linear (ReLU) activation, plus three neurons for the output layer.
We turn weights into fixed point integers and we quantize them to be computed on embedded hardware.
\\
\textbf{Reward function:} At each timestep, the agent receives a reward such as:
\begin{equation}
\label{eq:reward}
    r_t \triangleq 
    \begin{cases}
        1 - C*N_{TX}/N_{max},      & \text{if no losses}\\
        0,      & \text{otherwise}
    \end{cases}
\end{equation}
where $N_{TX}/N_{max}$ is the normalized retransmission factor and $C=\frac{3}{10}$ a constant controlling the efficiency-reliability tradeoff: low values favor high reliability, higher values encourage energy efficiency; and $N_{max}=8$ the maximum number of retransmissions achievable within a slot.


\textbf{Solvable RL problem.}
Due to space constraints, we give the intuition that Dimmer can be modeled as an RL problem.
Dimmer operates in periodic rounds.
Intuitively, the performance of the current round depends on the current interference as well as the chosen $N_{TX}$ and independent of past decisions.
Dimmer can then be modeled as an MDP, and is solvable using RL.

\textbf{Network-size independence.}
Neural networks suffer from their rigid structure: their input vectors must remain constant in size.
If a neural network is trained using one input per sensor node, the neural network needs to be entirely retrained if a node is added or removed.
To counter this limitation, our DQN requires input from a subset of nodes only, ordered from lowest reliability to highest.
Thus, Dimmer supports deployments of varying sizes without retraining for each new device.
We select the $K$ devices with lowest reliability to correctly represent the suffered packet losses.
Absence of feedback is treated as 0\% reliability, and 100\% radio-on time.

\textbf{Deep RL.}
We normalize inputs to $[-1,1]$.
We depict any reliability below 50\% -1, and 100\% reliability as 1.
For historical features, we represent the previous round as -1 if at least one packet was lost, and 1 if all packets were received by all nodes.
Thus, we obtain an input vector with 31 elements (see Table~\ref{tab:features}).
This enables a small neural network while being able to support a wide range of deployments with varying size.
Although we limit our input space, reliability and radio-on time are continuous values, and the state space is too large to use traditional Q-learning.
Instead, we must rely on deep Q-learning to solve Dimmer.

\textbf{Limiting the action space}.
Our reasons to use Decrease, Maintain, and Increase $N_{TX}$ rather than an action for each possible $N_{TX}$ values are two-fold:
(a) An action per value increases greatly the action space, thus causing a high resource overhead, i.e., a larger neural network, expensive in resource-constrained hardware, and extends the training time;
(b) In our experience, an action per value easily overfits the environment specifics, and behaves poorly against unseen dynamics.
Instead, a system with incremental updates learns that increasing prevents losses, irrespective of the current strategy.
A drawback is that the system is limited to step-wise increase, e.g., going from $N_{TX}=1$ to 4 takes three steps.

\begin{table}[tb]
     \caption{Input vector of Dimmer's DQN. Parentheses denote the number of elements used by Dimmer during evaluation.}
     \centering
     \begin{tabular}{ccc}
         \toprule
         Input & Number of rows (31) & Normalization\\
         \midrule
         Radio-on time & $K$ (10) & [0, 20ms] $\rightarrow$ [-1,1] \\
         Reliability & $K$ (10)& [50, 100\%] $\rightarrow$ [-1,1] \\
         $N$ parameter & $N_{max}+1$ (9)& One-hot encoding \\
         History & $M$ (2)& -1 if losses, otherwise 1 \\
         \bottomrule
         \vspace{0.1em}
     \end{tabular}
     \label{tab:features}
 \end{table}

\textbf{Trace environment.}
To train, an RL agent requires access to a physical deployment for hundreds of hours, or a simulated environment to speed up the learning process.
We create an environment from traces collected over multiple days, for different times of the day and frequencies.
We give the intuition on how traces are collected.
It is impossible to play out two actions ($N_{TX}+1$ and $-1$) with identical wireless conditions; we execute them sequentially, with minimal latency between.
Further, our jamming is controlled, so that all actions undergo similar conditions.
Since different actions are executed with minimal latencies, we approximate the effects of slow-fading interference.
Transient interference (in the order of ms), affects floods within a round, but not transitions, and is therefore correctly represented within our traces.

\textbf{Offline learning.}
Typical resource-constrained IoT platforms, with limited memory and CPUs, are unfit to train neural networks. 
Therefore, we train our neural network offline, and embed the result of the learning on the resource-constrained device for its inference step.
We train our DQN for 200~000 iterations with an epsilon-greedy selection scheme.
The selection probability in annealed from 100\% to 1\% linearly over the length of 100~000 steps, and fixed to a random action probability of 1\% afterward.
We select a discount factor $\gamma$ of 0.7.

\textbf{Embedded DQN.}
Our goal is a DQN able to run on various IoT platforms, even as resource-constrained as the old TelosB (4~MHz CPU, 10~kB RAM, no FPU).
Thus, we implement our own neuronal compute-system rather than use an existing framework, and use fixed-point integers for computation, set to 100 (two floating digits)~\cite{Lin2016-nn-quantization}.
By using 2B per weight and 4B for intermediary computation, our DQN uses 2.1~kB to store weights in flash, and 400~B of RAM for intermediary results.
On the old TelosB platform, a DQN execution takes 90~ms, due to 32-bit computation on a 16-bit CPU.
In comparison, a slot takes roughly 20~ms.
Thus, we execute the DQN after the last control slot.

\subsection{Distributed Forwarder Selection}
\label{sec:design:MAB}

\begin{figure}[tb]
    \centering
    \includegraphics[width=0.75\columnwidth]{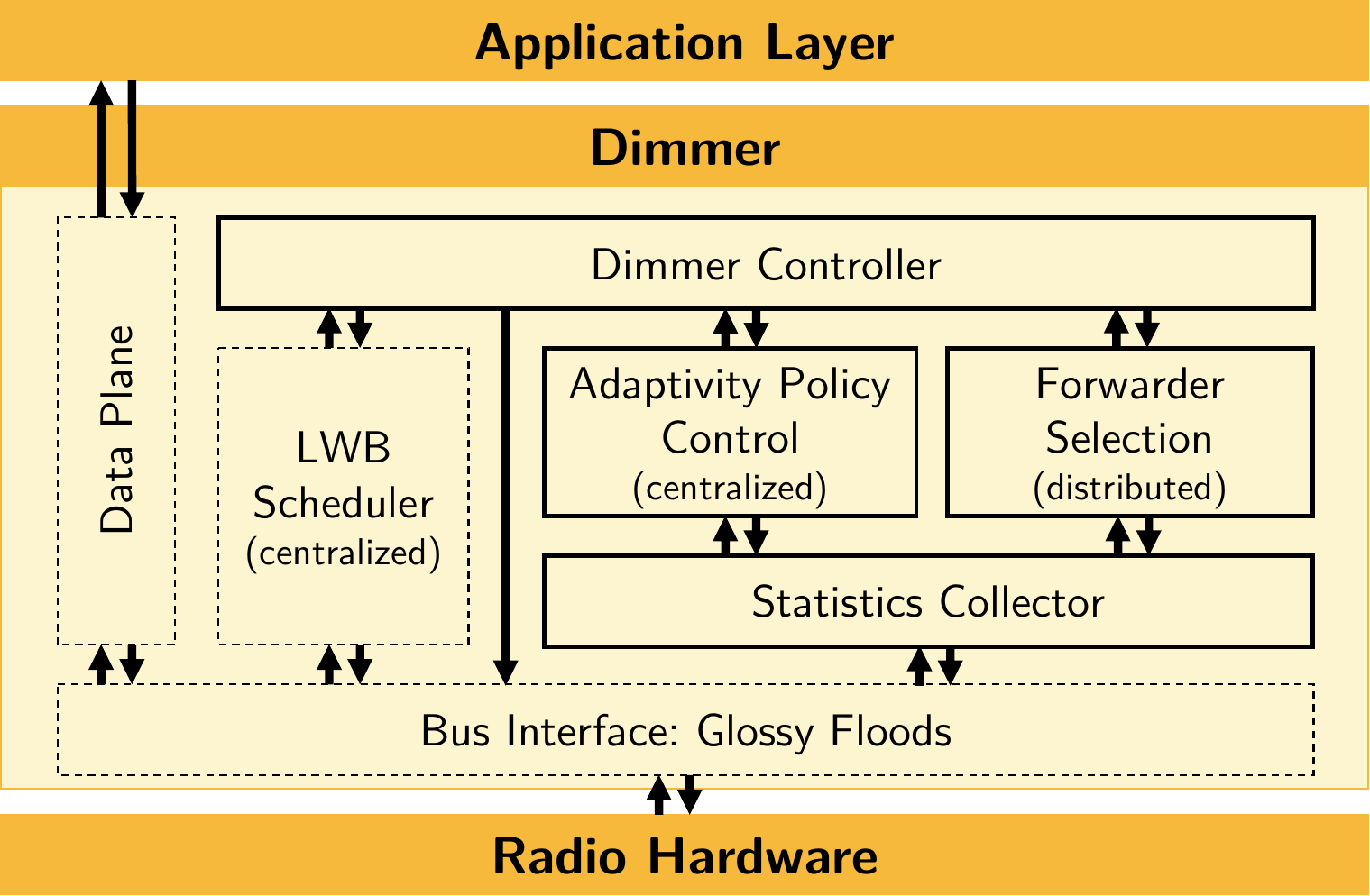}
    \caption{System architecture. 
        New components are highlighted in bold boxes.
        Adaptivity is centrally controlled via a deep Q-network, while forwarder selection is executed as a multi-armed bandits in a distributed fashion.}
    \label{fig:dimmer_arch}
\end{figure}

\textbf{Problem formulation.}
We formulate the problem of distributed forwarder selection as a two-armed bandit: each device locally decides whether it will act as an active forwarder (arm 1) or passive receiver (arm 2).
Specifically, we use Exp3~\cite{Auer2002-exp3}, and extend the concepts introduced for Glossy in~\cite{Zhang2017-LessIsMore}.
Here, we cannot use Q-learning: the number of input states is too large and building knowledge requires an unpractical amount of time.
Instead, we rely on fast exploration to find the optimal decision in the current state.
Further, from a local-device perspective, the environment is adversarial; distant devices' decisions affect our reward system.
Upper Confidence Bound (UCB) typically performs badly in adversarial environments, we must use an adversarial MAB, i.e., Exp3.
Sequentially, devices learn their role by randomly drawing a decision: if the communication experiences losses, we punish the chosen arm; if communication does not suffer losses, we reward the chosen arm.

\textbf{Improving stability.}
Due to the environment's non-stationarity, a distributed approach employing Exp3 is not guaranteed to converge: some devices might oscillate between active or passive, depending on precedent decisions.
We introduce three techniques to reduce the risk of oscillatory states that can degrade the network performance:
(a) Learning is sequential: each device has ten consecutive rounds to learn a role.
We thus improve the environment's stationarity for a given device.
(b) Network-breaking configurations are punished: Whenever a bad configuration is experienced, we reinitialize the passive arm to its initial value; thus greatly reducing the risk of re-entering this bad configuration.
(c) Learning follows a pseudo-random order: the learned configuration depends on the device order; devices that can try actions earlier are more likely to act as passive nodes.
The pseudo-random order ensures that learning is spread geographically, and that early passive receivers are not clustered together.

\subsection{System Architecture}
\label{sec:design:architecture}

We build Dimmer on LWB's 2019 reimplementation \cite{Mager2019}, and depict its architecture in Fig.~\ref{fig:dimmer_arch}.
Dimmer is composed of three main novel components: a \textit{statistics collector}, closing the feedback loop, the \textit{central adaptivity control} incorporating our deep-Q network (DQN), and a \textit{forwarder selection} implementing multi-armed bandits.
A controller manages and coordinates the different components, by updating internal scheduling parameters as well as the bus interface, and by polling the collected statistics.
The statistics collector has access to the LWB runtime and sent and received packets.

\textbf{Execution.}
Dimmer works in communication rounds, see Fig.~\ref{fig:central_adaptivity}.
A central coordinator starts the round with a control slot, during which it disseminates
both the communication schedule as well as an adaptivity command: a new global retransmission parameter $N_{TX}$, or a command allowing devices to execute locally their multi-armed bandits instances.
Immediately after the control slot, all nodes in the network apply the new $N_{TX}$ parameter.
Additionally, all devices measure their performance, i.e., reliability (packet reception rate, ratio of packets received / expected) and radio-on time at the end of each slot.
A series of data slots follow the control slot.
For each data slot, the source appends to its payload a two-byte header representing two performance metrics: its radio-on time averaged over the last floods, and its reliability (packet reception rate).
All receivers locally record the performance of distant devices.
Further, Dimmer uses slot-based channel-hopping; a static, global hopping-sequence is used for data slots, while all control slots are executed on channel 26.

\textbf{Global view.}
Dimmer continuously builds a global snapshot of the network, that is used both by the coordinator for adaptivity, and by other devices for the forwarder selection.
We estimate reliability via the schedule: if no message is received during an assigned slot, it is considered lost.
If no information is received from a given node, its feedback is locally filled with pessimistic values: 0\% reliability and 100\% radio-on time.

\subsection{Discussions}
\label{sec:design:discussion}

\textbf{Feedback latency.}
Devices share their local performance measured prior to their data slot, while the coordinator executes its DQN at the end of a round and shares the new value at the beginning of the next round.
If interference starts during a round, only the later transmitting devices can report its effects.
Thus, Dimmer takes one round to adapt to interference reported before the end of the round, or two rounds if more feedback is required. 
We note that although Dimmer's latency has no strong dependency on the number of nodes, the round periodicity might increase for large deployments.

\textbf{Scalability.}
Our DQN does not require feedback from all nodes: we use the K-lowest-reliability devices as input to the DQN.
Thus, Dimmer scales to deployments with varying numbers of devices without the need to change its architecture.
Similarly, Dimmer does not require all devices to provide their feedback; it is possible to define a subset of nodes that will not be accounted in the interference evaluation.
This, however, might leave some part of the network unprotected against localized interference.

\textbf{Centralized adaptivity.}
Interference near the coordinator can be harmful in centralized approaches.
However, like other LWB-class protocols, Dimmer requires nodes to receive the schedule packet to participate during the round.
In case a node missed a schedule slot, it will simply set its $N_{TX}$ to the global value once a schedule packet is received.

\textbf{MAB: long-term adaptivity.}
Contrary to the centralized DQN, our distributed MAB approach is slower to adapt.
$10\times D$ rounds are required to allow all $D$ devices to execute one learning iteration.
Further, the forwarder selection problem can be related to building a (non-spanning) tree; since there are many solutions to building a tree, our distributed selection scheme can oscillate between different configurations, and not converge to one.
This is by design, as we want Dimmer to adapt to changes in the topology, e.g., joining or leaving devices, and slow-fading links.

\textbf{Embedded DQN.}
In Dimmer, we do not assume the presence of an edge where computation can be offloaded.
For example, swarms of drones can not rely on constant connectivity to a server.
Instead, our DQN is embedded in our resource-constrained hardware.
\section{Evaluation}
\label{sec:evaluation}
\begin{figure*}[tb]
    \centering
    \begin{subfigure}[t]{0.26\textwidth}
        \vskip 0pt
        \captionsetup{width=.95\linewidth}
        \includegraphics[width=\textwidth]{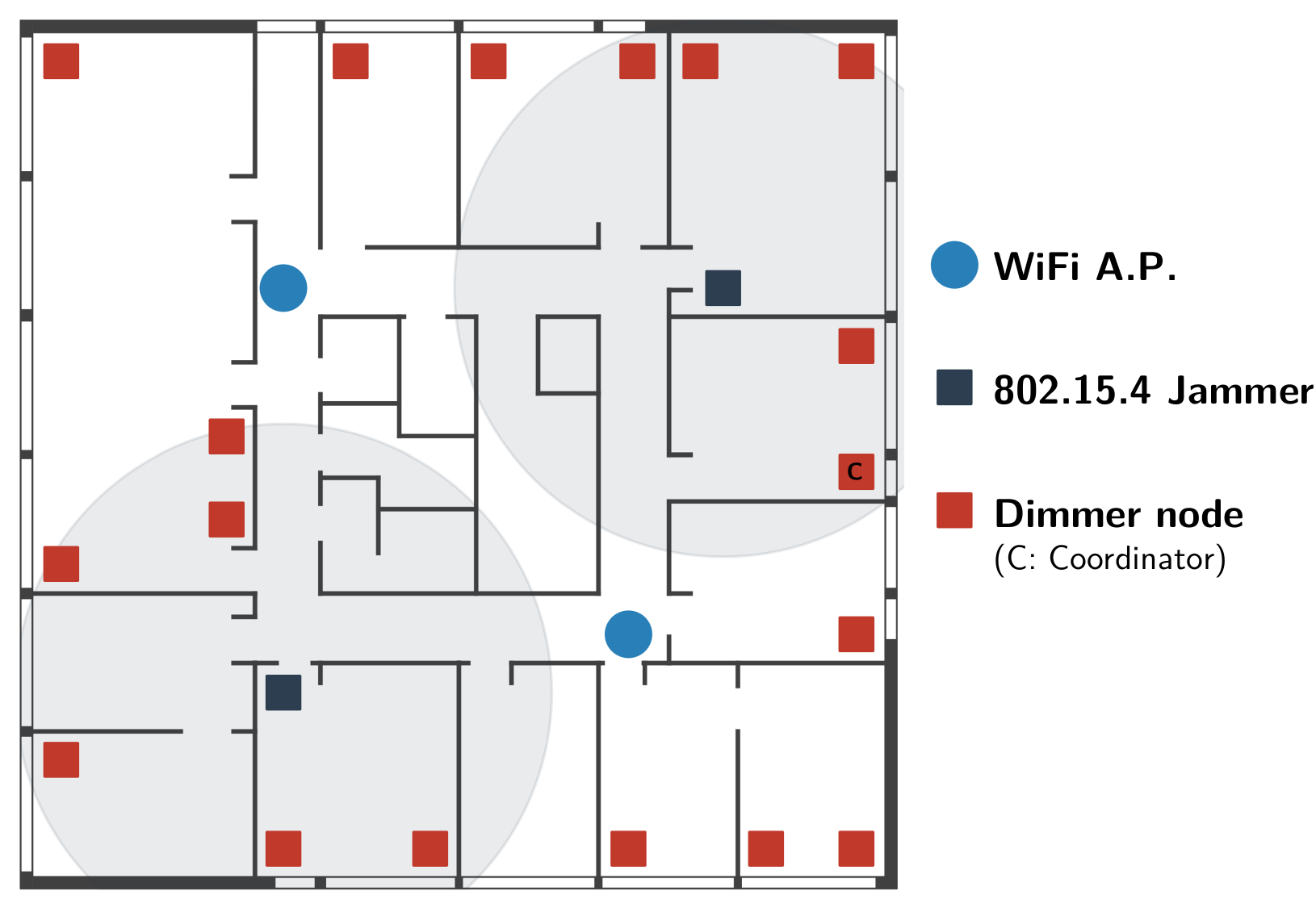}
        \caption{Testbed spanning $23\times23 m^2$. 18 TelosB compose a 3-hop network and share the medium with uncontrolled WiFi and Bluetooth PANs.
        Two additional TelosB act as 802.15.4 jammers. The central coordinator (C) is moderately affected by the nearest jammer.}
        \label{fig:testbed_map}
    \end{subfigure}
    \begin{subfigure}[t]{0.32\textwidth}
        \vskip 0pt
        \captionsetup{width=.9\linewidth}
        \includegraphics[width=\textwidth]{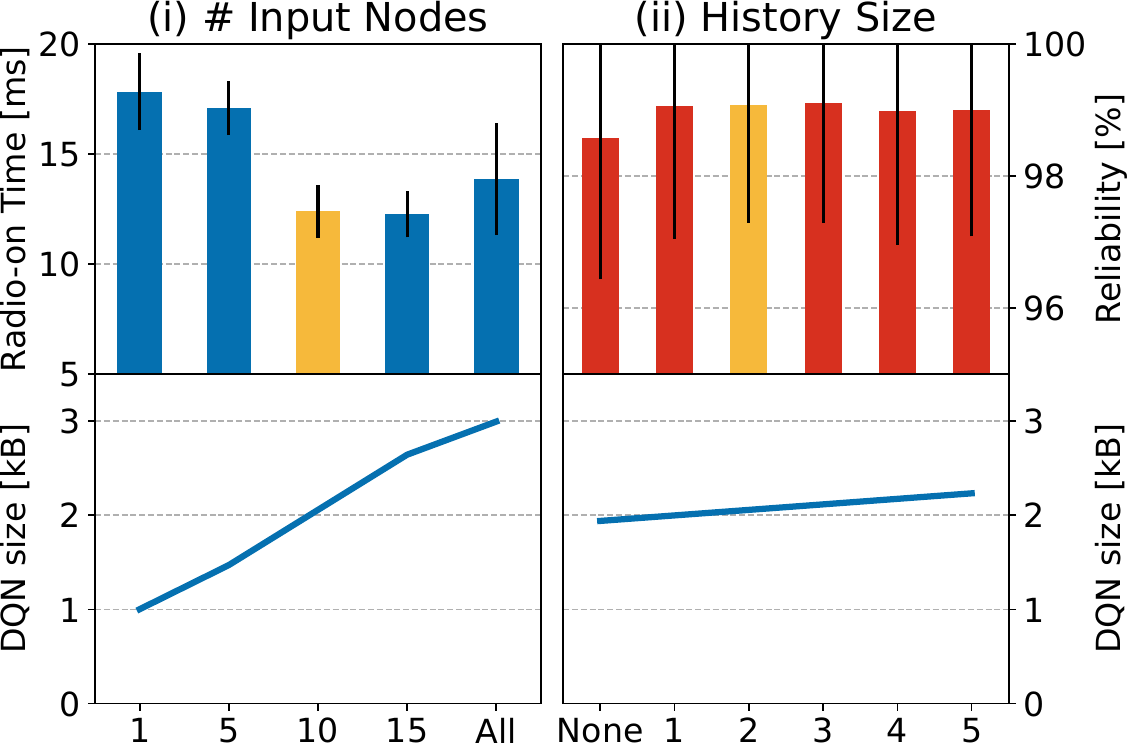}
        \caption{Input-feature selection.
        (i) Using only the worst node performance as input to the DQN leads to an overly conservative strategy that wastes energy but does not improve reliability.
        (ii) Using historical data improves reliability. We select 10 input nodes and 2 historical features.}
        \label{fig:eval_dqn_params}
    \end{subfigure}
    \begin{subfigure}[t]{0.36\textwidth}
        \vskip 0pt
        \begin{subfigure}[t]{\textwidth}
            \includegraphics[width=\textwidth]{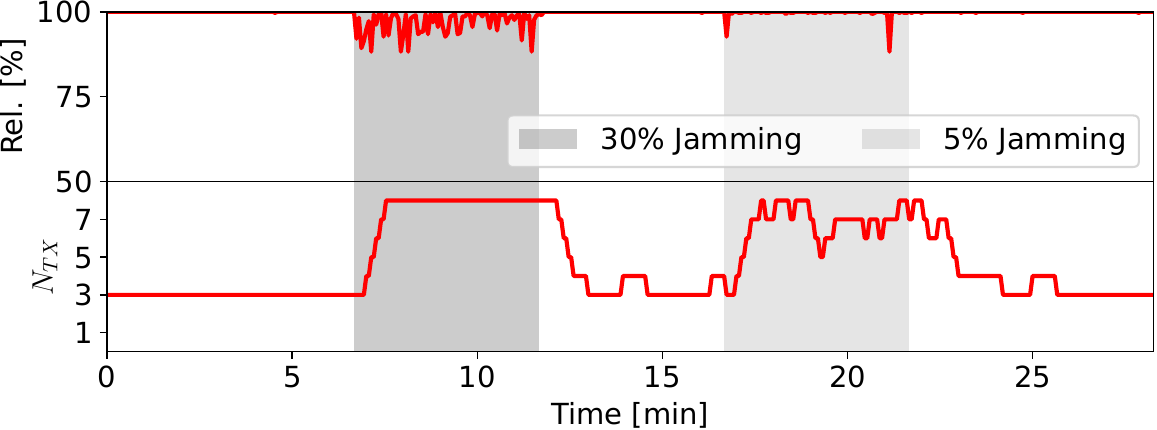}
            \caption{Dimmer. Reliability: 99.3\%, Radio-on time: 12.3~ms (average).}
            \label{fig:eval:adaptivity:dimmer}
        \end{subfigure}
        \begin{subfigure}[t]{\textwidth}
            \includegraphics[width=\textwidth]{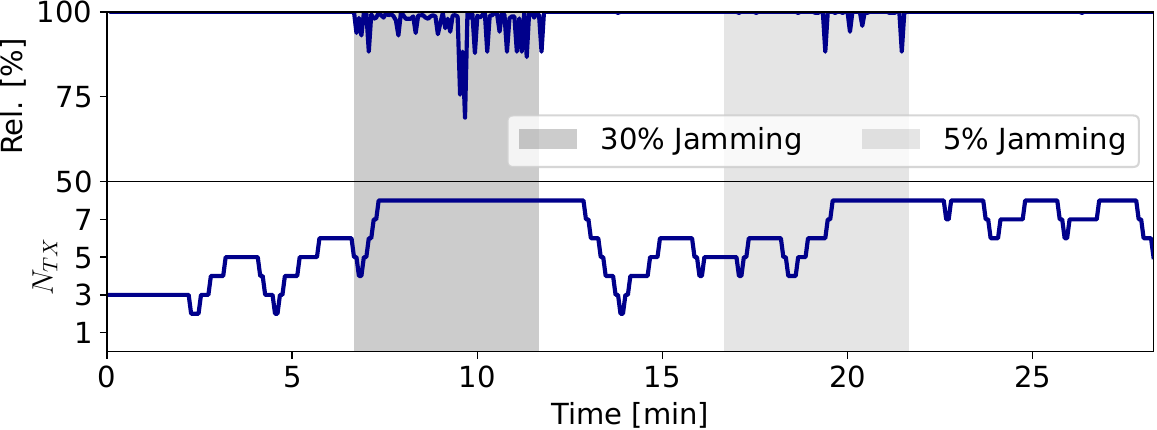}
            \caption{PID baseline. Reliability: 99.3\%, Radio-on time: 14.4~ms (average).}
            \label{fig:eval:adaptivity:PID}
        \end{subfigure}
    \end{subfigure}
    \caption[caption]{Tuning and evaluation Dimmer.}
    \label{fig:eval:adaptivity}
\end{figure*}

We evaluate Dimmer on two testbeds of respectively 18 and 48 TelosB devices.
We lay out our methodology in \S\ref{sec:eval:setup} and investigate in \S\ref{sec:eval:dqn} how input features affect Dimmer's performance.
Then, we quantify in \S\ref{sec:eval:adaptivity} adaptivity to dynamic interference.
We evaluate in \S\ref{sec:eval:mab} our forwarder selection approach, and finally investigate how our DQN behaves on new topologies without retraining in \S\ref{sec:eval:graz}.

\subsection{Setup and Methodology}
\label{sec:eval:setup}

\textbf{Implementation.}
We implement Dimmer for Contiki-NG, based on LWB's 2019 reimplementation~\cite{Mager2019}.
Dimmer is hardware agnostic; we use the TelosB platform (4~MHz 16-bit CPU, 10 KB of RAM, 48~KB of firmware storage, CC2420 radio for IEEE~802.15.4) throughout this evaluation.
Dimmer and its training environment are open-source (see \S\ref{sec:introduction}).

\textbf{Testbeds.}
We evaluate Dimmer on two testbeds:
(1) Our 18-device, 3-hop deployment, see Fig.~\ref{fig:testbed_map}.
Located in offices and lab rooms, it shares the spectrum with WiFi and many Bluetooth PANs (from cellphones, headphones, etc.), all outside of our control, and create interference during work hours.
We further use two additional TelosB as jammers, and inject controlled IEEE 802.15.4 interference using Jamlab~\cite{Boano2011-Jamlab}.
The central-coordinator's reception is (moderately) perturbed by the nearest jammer.
\\
(2) In \S\ref{sec:eval:graz}, we evaluate Dimmer on the public testbed D-Cube, featuring 48 TelosB motes and controlled WiFi interference~\cite{Schuss2017-DCube}.
Interferer locations and topology are unknown to us, coordinator is device ID 202.

\textbf{Baselines.}
To evaluate the effectiveness of deep RL, we implement and tune a PI controller as adaptive baseline.
PIDs are among the most common solutions for closed-loop systems~\cite{Ang2005-PID-tuning}, and make for a good comparison for traditional vs RL methods.
We set $K_P=1$ and $K_I= 0.25$.
We tune the baseline PI controller through experiments on the deployment, to maximize reliability first, and minimize energy consumption if reliability is at 100\%.

We evaluate against Crystal in \S\ref{sec:eval:graz}, a dependable ST protocol for aperiodic collection~\cite{Istomin2018-Crystal2}.
Crystal uses channel-hopping, acknowledgements, and noise detection to achieve high performance in the presence of strong interference (see \S\ref{sec:related_work}).
We use the parameters provided by the designers for the EWSN 2019 dependability competition~\cite{Trobinger2019-Crystal-comp}, i.e., the exact scenario evaluated here.
While comparing against Crystal, we activate channel-hopping and packet acknowledgements (a message is sent again if no ACK is received) in Dimmer.

\textbf{Interference scenarios.}
We run the following scenarios:
\begin{itemize}
    \setlength\itemsep{0em}
    \item No interference: experiments run at night on channel 26, no injected interference.
    \item Controlled 802.15.4 interference: We use the CC2420 radio of two additional TelosB jammers (see Fig.~\ref{fig:testbed_map}) to jam communication at 0~dBm on channel 26.
    We jam communication with 13~ms TX bursts, which corresponds to a typical WiFi burst of packets \cite{Boano2011-Jamlab}.
    Bursts are periodically repeated; a 10\% interference corresponds to a 13~ms burst every 130~ms, a 35\% interference ratio represents a 13~ms burst every 37~ms. In comparison, a Glossy flood is given at most 20~ms to communicate.
    \item D-Cube~\cite{Schuss2017-DCube}: We use the public testbed D-Cube, featuring controlled WiFi interference. Experiments run at night, using channel-hopping.
\end{itemize}

\textbf{Metrics.}
We evaluate the following metrics:
\begin{itemize}
    \item Radio-on time: the amount of time the radio has been active (i.e., listening or transmitting) for one slot, averaged over all slots. Slots in which no packet was received are accounted for. 
    \item Reliability: the percentage of destinations that correctly receive a packet.
\end{itemize}

\textbf{Parameters.}
We list the network parameters used throughout our evaluation, respectively on our testbed and in D-Cube:
(a) Rounds have a period of 4~sec, 1~sec in D-Cube.
(b) Slots have a maximum duration of 20~ms.
(c) We use periodic 4-sec broadcast traffic from all 18 devices in our testbed, i.e., we require 18 data-slots. We use 10 source-nodes with 1-sec traffic period in D-Cube. 
(d) Packets are 30~B long, including 3-byte LWB and 2-byte Dimmer headers.
(e) Dimmer transmits at 0~dBm.

\subsection{Deep-Q Network Features Selection}
\label{sec:eval:dqn}

We investigate how the number of inputs, i.e., the amount of devices' feedback and historical features affect the behavior of Dimmer.
We collect an evaluation dataset of 25~000 samples over channel 26, featuring some periods of mild and heavy interference, and some interference-free episodes.
For each parameter we evaluate, we train three models, and average the overall performance over those models.
For each model, we run 100 episodes comprising 100 consecutive decisions each for the number of nodes and reward.
Throughout the section, error bars represent standard deviation.

\textbf{Number of devices.}
We evaluate how many devices' inputs are necessary for Dimmer.
As described in \S\ref{sec:design:DQN}, Dimmer selects $K$ nodes with lowest-reliability as input to its DQN.
We fix the number of historical features to 2, and vary $K$ from one, i.e., only the device with lowest reliability is used, up to all 18 devices used as input.
\\
\textbf{Results.}
Fig.~\ref{fig:eval_dqn_params}(i) depicts the effect of $K$ on Dimmer's radio-on time.
Limited device subsets ($K=$ 1 to 5) lead to conservative policies, with high retransmission parameters even under non-interfered episodes, thus wasting energy.
Using all 18 devices, the DQN overfits the deployment and reacts to transient losses, thus also wasting energy after short-term interference.
Note that reliability is roughly constant for all experiments, i.e., conservative policies do not provide higher protection against losses on average.
For the remainder of this evaluation, we choose $K=10$ as our input, which both minimizes the radio-on time, as well as provides a good trade-off w.r.t. the neural network size.

\textbf{History size.}
We evaluate how historical features affect the DQN results.
We focus on short decision updates in low and mild-interference episodes, to test Dimmer's ability to distinguish transient disturbances from longer-term interference.
We average the results over 1000 episodes of 2 consecutive decisions.
\\
\textbf{Results.}
Fig.~\ref{fig:eval_dqn_params}(ii) shows the impact of historical features.
Adding historical data helps Dimmer differentiate transient interference that affects a single round, from long-term interference that must be dealt with.
In the absence of historical features, the DQN obtains 98.5\% reliability on average, while it achieves 99\% with historical features.
Adding further historical features does not seem to have a measurable impact on the overall performance in our evaluation.
For the remainder of this evaluation, Dimmer uses two historical features (i.e., data about losses over the last 8~sec).

\subsection{Adaptivity Against Interference}
\label{sec:eval:adaptivity}

\begin{figure*}[tb]
    \centering
    \begin{subfigure}[t]{0.48\textwidth}
        \vskip 0pt
        \includegraphics[width=\textwidth]{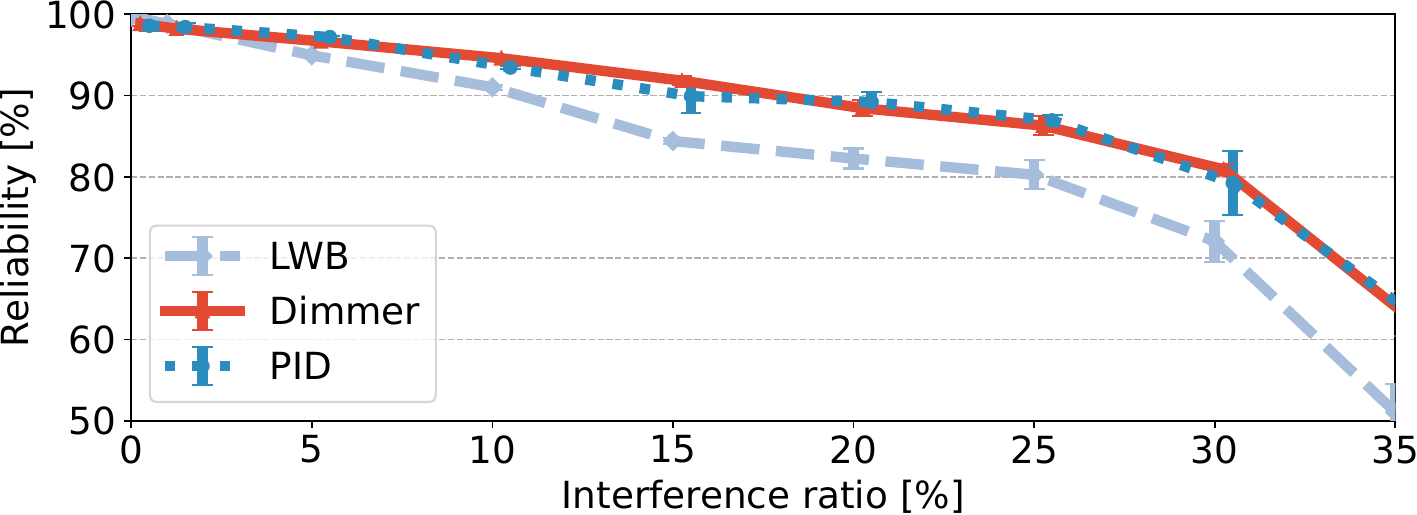}
        \caption{Reliability against interference. Both Dimmer and the PID baseline tackle interference with increased retransmissions. Because some slots manage to fit between two interference bursts, LWB is able to maintain some communication.}
        \label{fig:eval:reliability-vs-interf}
    \end{subfigure}
    \hfill
    \begin{subfigure}[t]{0.48\textwidth}
        \vskip 0pt
        \includegraphics[width=\textwidth]{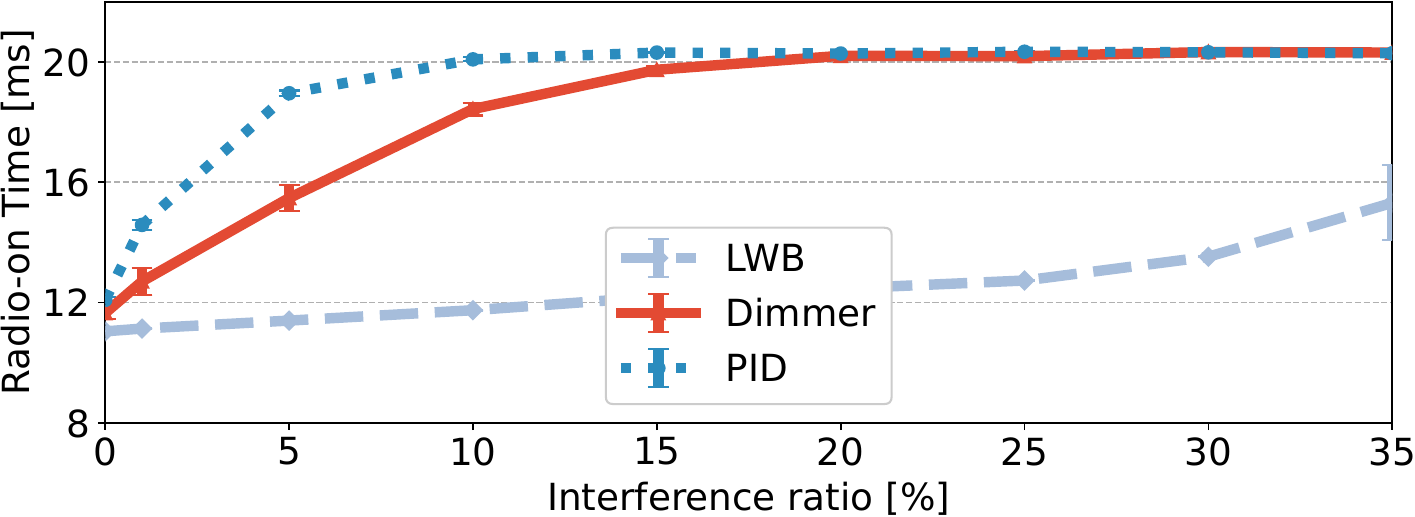}
        \vskip 0pt
        \caption{Radio-on time against interference. The PI controller quickly reacts to interference and over-provisions retransmissions, leading to a maximized energy consumption. In contrast, Dimmer adapts to the interference level, leading to a lower radio-on time compared to the PID, for similar  reliability.}
        \label{fig:eval:radio-on-vs-interf}
    \end{subfigure}
    \caption[caption]{Adaptivity to intermediate interference levels.}
    \label{fig:eval:vs-interf}
\end{figure*}
Next, we evaluate how Dimmer and the PID baseline adapt to interference.
First, we investigate adaptivity against dynamic interference.
Then, we evaluate reliability and radio-on time against static levels of disturbances.

\textbf{Dynamic interference.}
We operate Dimmer and the PID baseline on our deployement on channel 26, during the day (see \S\ref{sec:eval:setup}, Fig.~\ref{fig:testbed_map}).
We use two additional TelosB devices to inject IEEE 802.15.4 interference.
The experiment starts with the jammers off.
After 7~min, we inject heavy interference occupying the medium 30\% of the time (a 13~ms burst at 0~dBm, repeated every 43~ms).
The interference lasts 5~min, after which we turn off the jammers.
After 5~min of calm, we inject  light interference, occupying the medium 5\% of the time (13~ms burst at 0~dBm, every 230~ms). 
After 5~min, the jammers are turned off.

\textbf{Results.}
Fig.~\ref{fig:eval:adaptivity:dimmer} depicts a typical execution of Dimmer in the presence of dynamic interference, while Fig.~\ref{fig:eval:adaptivity:PID} depicts the PID baseline against the same scenario.
In the absence of interference, Dimmer learns that $N_{TX}=3$ provides high reliability.
The PID baseline slowly reduces $N_{TX}$ to save energy, and increases if losses are experienced; it oscillates around $N_{TX}=3$ in the absence of injected disturbances.
Both Dimmer and the PID baseline detect the heavy interference (30\%) we inject, and react by increasing the number of retransmissions.
Both protocols also decrease once the interference has stopped, although the PID baseline is slower here due to its integrative component.
Under light interference (5\%), Dimmer detects that allowing the maximum amount of retransmissions is not necessary, and searches for the optimal setpoint for the current disturbance.
The PID baseline, after detecting losses, overshoots to the maximum retransmission; due to its I component, it converges slowly back to normal.
\\
Both Dimmer and the PID baseline provide 99.3\% reliability over the experiment, yet Dimmer requires only 12.3~ms of radio-on time, while the traditional PID baseline, oscillating during interference-free and overshooting under interference, requires 14.4~ms.

\textbf{Interference levels.}
We investigate how Dimmer and the PI baseline scale to various interference levels, to see whether Dimmer overshoots interference or is able to measure interference strength.
Comparison against state-of-the-art protocols is carried in \S\ref{sec:eval:graz}.
We run Dimmer, the adaptive PID baseline, and the static LWB ($N_{TX}=3$) against a continuous, static interference-pattern, ranging from 0\% (no interference) to 35\% (13~ms burst every 37~ms).
Results are averaged over all rounds of three 30-minute runs, for each interference level; error bars represent standard deviation between rounds.

\textbf{Results.}
Fig.~\ref{fig:eval:reliability-vs-interf} depicts the reliabilty of Dimmer, LWB and the PI baseline, Fig.~\ref{fig:eval:radio-on-vs-interf} the average time their radios were active per slot, a proxy for energy consumption.
As interference arises, reliability of all protocols decreases; both the PID baseline and Dimmer allow communication to survive to higher interference levels.
Since the PID baseline is unable to quantify interference levels, radio-on time quickly grows to the maximum slot size, 20~ms.
Dimmer is able to distinguish interference strength, and requires less energy than the PID baseline for low interference (below 15\%, 13~ms burst every 87~ms), for similar reliability.
At higher interference strength, all retransmissions are necessary, Dimmer requires 20~ms radio-on time.
As interference arises, LWB requires more time to achieve its 3 static receptions and retransmissions within a flood, but never requires the full slot duration on average.

\textbf{Main findings.}
Both the PID baseline and Dimmer are adaptive: interference is detected and counteracted through higher retransmissions.
Yet, the PID baseline is unable to distinguish interference levels, overshoots under low interference, thus wasting energy.
More complex PID systems are possible, yet they require expert knowledge to be designed and tuned.
In contrast, Dimmer distinguishes interference strength, and does not require
expert knowledge: from unlabelled traces, it learns how to deal with interference patterns.

\subsection{Forwarder Selection with MAB}
\label{sec:eval:mab}
\begin{figure}[tb]
    \centering
    \includegraphics[width=0.8\columnwidth]{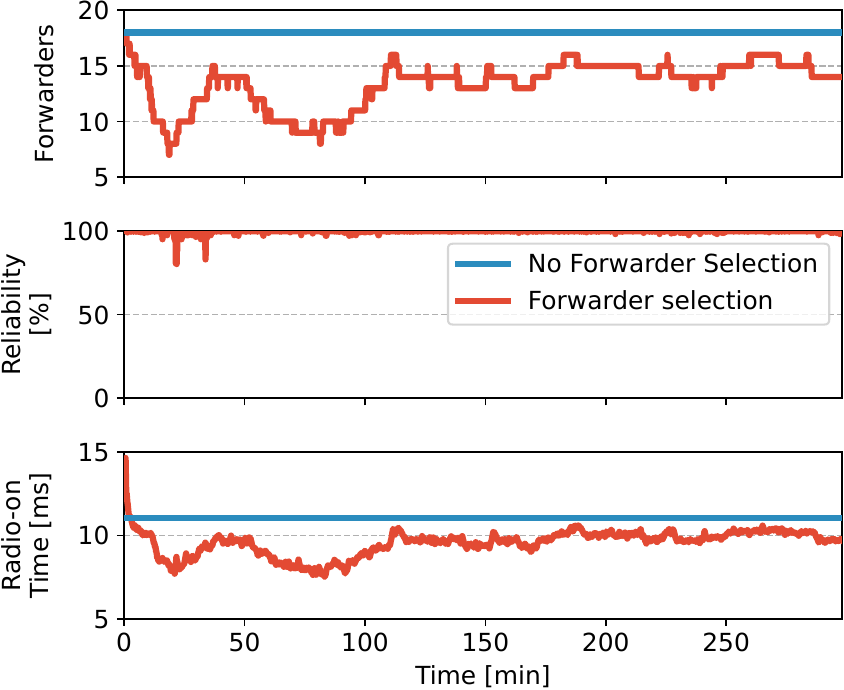}
    \caption{Forwarder Selection with Multi-Armed Bandits. Nodes takes turn learning whether to act as forwarder. As learning occurs, the number of forwarders decreases, but learning might cause transient packet losses.}
    \label{fig:eval:mab}
\end{figure}
Next, we evaluate how Dimmer deactivates superfluous transmitters during interference-free episodes.

\textbf{Scenario.}
We execute the forwarder selection scheme on channel 26 during the night, for 5~hours.
During that time, the DQN is deactivated.
We show that the forwarder selection alone prohibits breaking configurations.
Each of the 18 devices is sequentially given 10 consecutive rounds to learn a decision (1) act as forwarder or (2) act as passive receiver.
A single 5-hour learning instance is depicted.

\textbf{Results.}
Fig.~\ref{fig:eval:mab} shows (a) the number of active forwarders, (b) the reliability, and (c) the average radio-on time as a function of time.
During the first two hours, we see a rapid decrease of active forwarders; devices are encouraged to try passivity.
At the 30~min mark, the first network-breaking configuration is encountered; the learning passive devices are punished and reliability is maintained.
A similar event happens after the first-hour mark.
Then, devices maintain a conservative configuration, with around 14 active forwarders and 4 passive receivers.
As learning is continuous, devices tend to exchange roles, acting passive for a while, and then helping the dissemination as forwarder.
Over the five hours, Dimmer achieves a reliability of 99.9\%; for an average radio-on time of 9.55~ms (against 11.04~ms in the absence of forwarder selection).

\textbf{Main findings.}
By favoring conservative decisions, our forwarder selection prohibits network-breaking configurations.
Further, Dimmer continuously adapts to the slow-fading of its environment.

\subsection{Performance on Unknown Deployments}
\label{sec:eval:graz}
\begin{figure}[tb]
    \centering
    \includegraphics[width=\columnwidth]{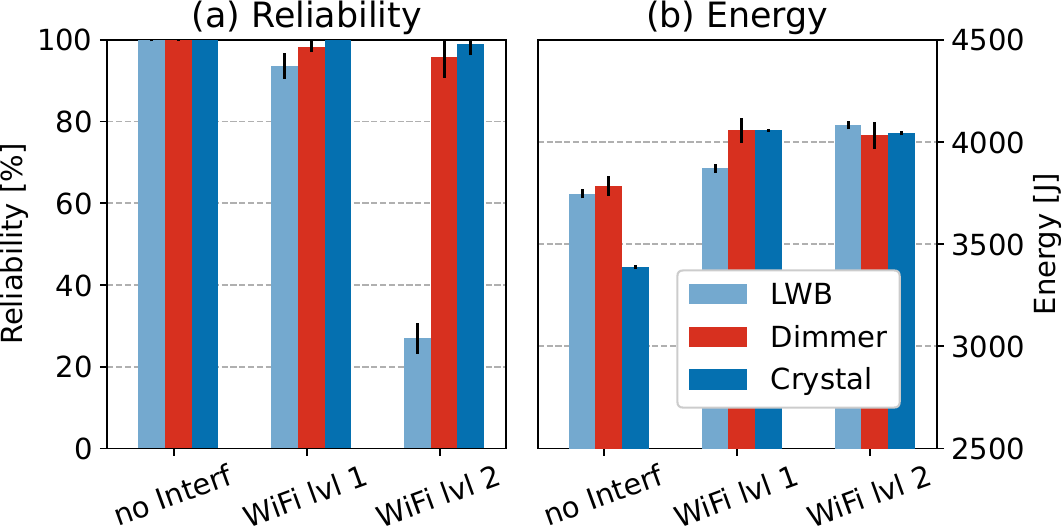}
    \caption{Dimmer on the 48-device D-Cube. Without retraining, Dimmer is able to tackle previously unseen 802.15.4 and WiFi patterns.}
    \label{fig:eval_graz}
\end{figure}

Next, we evaluate how Dimmer performs on a larger deployment, for which it has not trained for, against unseen WiFi interference.
Further, we investigate how Dimmer compares to the state-of-the-art with Crystal~\cite{Istomin2018-Crystal2}.

\textbf{Scenario.}
We execute Dimmer on D-Cube, featuring 48 TelosB devices forming an unknown topology~\cite{Schuss2017-DCube}.
We evaluate an aperiodic data collection scenario (Data Collection V1, used in the EWSN competitions): five known sources transmit packets at random intervals to a known sink; reliability represents the number of packets received at the sink.
We compare Dimmer with the baseline LWB ($N_{TX}=3$), as well as Crystal, a dependable ST protocol for aperiodic collection~\cite{Istomin2018-Crystal2}.
Crystal has been configured after preliminary trials on the deployment, while our DQN has not been trained for this topology. Interference patterns are unknown to both protocols.
Here, we execute Dimmer with channel-hopping and packet acknowledgements. 
We run three episodes: (a) interference-free, (b) WiFi interference (level 1), and (c) WiFi interference (level 2), where the WiFi levels are defined by the D-Cube maintainers.
Results are averaged over ten 10-minute experiments, error bars denote standard deviation.

\textbf{Results.}
Fig.~\ref{fig:eval_graz} depicts the reliability and energy consumption of LWB, Dimmer, and Crystal.
LWB is single-channel and best-effort, all packets are received in the absence of interference, but reliability drops to 93.6\% and 27\% under WiFi interference, while energy increases due to higher failed receptions and lost synchronization.
Crystal's design relies on expert knowledge and its configuration on preliminary trials; all packets are received against the first WiFi interference level, while 99\% are received against the second level.
With Dimmer, we do not rely on expert knowledge, nor did we collect traces for this specific testbed and interference patterns.
We reuse the DQN trained for 18-nodes against 802.15.4 interference, and simply add application-layer ACKs.
Dimmer achieves 100\% reliability in the absence of interference, and 98.3\% and 95.8\% against WiFi levels 1 and 2.
Compared to LWB, energy usage greatly increases as soon as interference is detected as Dimmer increases the number of retransmission to $N_{TX}=8$, yet energy consumption is comparable to the state-of-the-art Crystal.
As interference is much stronger than evaluated before, the PID baseline provides similar performance as the DQN under interference, but oscillates heavily under normal conditions, and is not depicted here.

\textbf{Main findings.}
Without retraining, Dimmer performs on a new topology with $2.6\times$ the number of devices, and is able to react to unseen WiFi interference patterns.
Further, without expert knowledge and without traces collected from the new deployment, Dimmer approaches the performance offered by the state-of-the-art Crystal protocol.
\section{Related Work}
\label{sec:related_work}

\textbf{AI-enabled wireless.}
Machine learning has been used in the literature to tackle the problem of interference identification.
Using SVM, SoNIC classifies different interference patterns (WiFi, Bluetooth, etc.), using features such as RSSI and error burst spanning~\cite{Hermans2013-Sonic}.
Grimaldi~et~al. rely on classification trees and multiclass SVM to provide interference classification~\cite{Grimaldi2019-IDI}.
In both approaches, classification is separated from the system reaction to interference, and requires an annotated dataset.
Instead, RL provides tools to learn how to directly operate over the wireless medium.
Amuru~et~al.~use post-decision state learning to learn how to backoff in CSMA/CA~\cite{Amuru2015-RL-CSMA}.
Mastronade~et~al.~also employ post-decision state learning and propose a rate-adaptive flavor of CSMA/CA~\cite{Mastronarde2016-RL-CSMA}.
Zhu~et~al.~rely on deep Q-learning to schedule transmissions on relay devices~\cite{Zhu2017}.
Dakdouk~et~al.~propose to use MABs and the Upper Confidence Bound to select the next channel in 802.15.4 TSCH~\cite{Dakdouk2018-RLTSCH}.
In Less is More, Zhang~et~a.~use Exp3 to optimize individual-device retransmission at runtime in Glossy~\cite{Zhang2017-LessIsMore}.
Their work differs from ours on two major points:
(a) They investigate optimizing Glossy (one-to-all) in the interference-free case; our forwarder selection solves a larger problem generalized to wireless-bus abstractions (all-to-all). 
(b) They require a feedback header that changes during a flood, thus breaking constructive interference. Our solution provides feedback with the next round, thus maintaining Glossy's properties.

\textbf{Dependable ST protocols.}
A large body of literature studies dependability in ST protocols.
Robust Flooding improves over Glossy through TX-based channel-hopping and additional retransmissions~\cite{Lim2017-RobustGlossy}.
Al Nahas et al.~also rely on channel-hopping for A\textsuperscript{2}~\cite{AlNahas2017-A2}.
DeCoT+, relies, among other things, on channel-hopping, payload retransmissions, and network-coding to survive harsh interference~\cite{Ma2019-DeCOT}.

Istomin et al.~improve Crystal against heavy interference~\cite{Istomin2018-Crystal2}.
In addition to the original Transmission-Acknowledgement (TA) scheme, allowing Crystal to survive against collisions and transient losses, the authors extend their original design with TA-pair channel-hopping and noise detection.
If noise is detected, additional TA pairs are available before turning off the radio.
Crystal is adaptive: upon noise detection, additional slots are added.
Yet, those general parameters were obtained via expert knowledge.
In contrast, Dimmer is self-adaptive: we do not retrain the DQN when changing from periodic to aperiodic collection and from 802.15.4 interference to WiFi, nor when we evaluate on a 48-device testbed featuring strong WiFi jamming, after training with 18 devices and 802.15.4 jamming.
\section{Conclusion}
\label{sec:conclusion}

Synchronous Transmissions (ST) are a high-performance and energy-efficient communication paradigm in low-power wireless networks.
Through the use of custom rules, hand-tailored parameters, and additional  retransmissions, dependable ST are shown to provide high-performance in harsh and highly-interfered environments.
Yet, tuning those dependable ST protocols is arduous, requiring expert knowledge, extensive testing, and is often achieved for a specific deployment and given scenario.
We introduce Dimmer, a self-adaptive ST protocol.
Through the use of deep reinforcement learning, from unlabeled traces, and in the absence of human supervision, Dimmer learns by itself how to detect interference and how to adapt its retransmission parameter to maintain communication in harsh and dynamic environments.
Further, using multi-armed bandits, Dimmer deactivates superfluous transmitters at runtime, thus saving energy and adapting to the long-term dynamics of its deployment.
We show that Dimmer achieves 95.8\% reliability against strong WiFi interference, thus approaching the performance of the state-of-the-art Crystal, in the absence of any human supervision.
Further, Dimmer is able to support new deployments and unseen interference without retraining.

\section*{Acknowledgement}
We would like to thank Gian Pietro Picco, Matteo Trobinger and Timofei Istomin for their invaluable input on Crystal.

\bibliographystyle{IEEEtran}
\bibliography{IEEEabrv,biblio}

\end{document}